\documentclass[12pt]{iopart}

\usepackage{iopams}
\usepackage{graphicx}  \usepackage{dcolumn}  \usepackage{bm} \usepackage{epsfig} \usepackage{fancyhdr}  \usepackage{enumerate}  \usepackage{float} \usepackage{setspace}  \usepackage{fancybox}
\usepackage{verbatim}
\usepackage{braket}

\def\ii{{\rm i}}

\begin{document}

\title{Second-order quantum nonlinear optical processes in single graphene nanostructures and arrays}
\author{M T Manzoni$^1$, I Silveiro$^1$, F J Garc\'{\i}a de Abajo$^1$$^2$ and D E Chang$^1$}
\address{1 ICFO - The Institute of Photonic Sciences, Mediterranean Technology Park, 08860 Castelldefels (Barcelona), Spain}
\address{2 ICREA - Instituci\'o Catalana de Recerca i Estudis Avan\c{c}ats, Passeig Llu\'{\i}s Companys 23, 08010 Barcelona, Spain}
\eads{\mailto{darrick.chang@icfo.es}}

\date{\today}

\begin{abstract}
Intense efforts have been made in recent years to realize nonlinear optical interactions at the single-photon level.\,\,Much of this work has focused on achieving strong third-order nonlinearities, such as by using single atoms or other quantum emitters, while the possibility of achieving strong second-order nonlinearities remains unexplored. Here, we describe a novel technique to realize such nonlinearities using graphene, exploiting the strong per-photon fields associated with tightly confined graphene plasmons in combination with spatially nonlocal nonlinear optical interactions. We show that in properly designed graphene nanostructures, these conditions enable extremely strong internal down-conversion between a single quantized plasmon and an entangled plasmon pair, or the reverse process of second harmonic generation. A separate issue is how such strong internal nonlinearities can be observed, given the nominally weak coupling between these plasmon resonances and free-space radiative fields. On one hand, by using the collective coupling to radiation of nanostructure arrays, we show that the internal nonlinearities can manifest themselves as efficient frequency conversion of radiative fields at extremely low input powers. On the other hand, the development of techniques to efficiently couple to single nanostructures would allow these nonlinear processes to occur at the level of single input photons.

\end{abstract}
\submitto{\NJP}
\maketitle


The ability to realize strong interactions between single photons potentially enables one to attain the ultimate limit of classical nonlinear optical devices \cite{CSD07,FOS13,TGP13} and is a key resource in quantum information processing \cite{K08}. A number of schemes are being pursued to realize third-order nonlinearities at the quantum level \cite{rabl11,NLO_Rydberg13,peyronel12}, most notably by exploiting the anharmonic electronic spectrum associated with individual atoms or other quantum emitters \cite{Turchette95,BBM05,Fushman08,RS14,Tey08,Chang06}. However, there still exist no viable approaches toward achieving comparably strong second-order nonlinearities. For example, in state-of-the-art devices, a single photon is down-converted into an entangled photon pair with a relatively low efficiency of $\sim$\,$10^{-8}$ \cite{SAN11,GUE14}. Devices with higher efficiencies would be useful for many significant tasks, such as dramatically improving the fidelities of quantum information processing schemes based upon detection and post-selection \cite{SAN11}.

In this Letter, we show that graphene is a promising second-order nonlinear material at the single-photon level due to its extraordinary electronic and optical properties \cite{CGP09}. This approach makes use of the fact that a conductor enables a nonlinear optical interaction that is spatially nonlocal over a distance comparable to the inverse of the Fermi momentum $k_F$. In graphene, this length can be electrostatically tuned to be significantly larger than in typical conductors. At the same time, graphene can support tightly confined surface plasmons (SPs) --combined excitations of electromagnetic field and charge density waves-- whose wavelength is reduced well below the free-space diffraction limit \cite{paper176} and whose momentum $q_p$ is consequently enlarged. We show that the ability to achieve ratios $q_p/k_F$ approaching unity enables giant second-order interactions between graphene plasmons.

We first study the implications of such nonlinearities in a finite-size nanostructure, obtaining a general scaling law for the nonlinearity as a function of the linear dimension of the structure and the doping. To give an explicit example, we compute numerically the nonlinearities associated with a structure designed to support plasmon resonances at frequencies $\omega_p$ and $2\omega_p$, which enables second harmonic generation (SHG) or down conversion (DC). Under realistic conditions, we find that the rate of internal conversion between a single quantized plasmon in the upper mode and two in the lower mode can be roughly $1\%$  of the bare frequency, indicating a remarkable interaction strength.

It is not straightforward to directly observe plasmons, and instead they are typically excited and coupled out to propagating photons with low efficiencies. Thus, we then investigate how the extremely strong internal nonlinearities can manifest themselves given free-space input and output fields. First, we show that the collectively enhanced coupling of an array of nanostructures to free-space fields enables an extremely low-intensity input beam to be converted to an outgoing beam at the second harmonic, via interaction with plasmons. Next, we derive an important fundamental result, that while such an array can collectively increase the linear coupling between free fields and plasmons, it ultimately $dilutes$ the effect that the intrinsic nonlinearities of plasmons can have on these free fields. Motivated by this, we finally argue that it is crucial to develop techniques to couple efficiently to single nanostructures. We show that efficient coupling would enable SHG or DC with inputs at the single-photon level, and  predict a set of experimental signatures in the output fields that would verify that strong quantum nonlinear interactions are occurring between graphene plasmons. 

\section{Second-order nonlinear conductivity of graphene}

Graphene has attracted tremendous interest due to its ability to support tightly confined, electrostatically tunable SPs \cite{paper176,HD07,WSS06,JBS09,paper196,FRA12,paper212,BJS13}. More recently, the nonlinear properties have gained attention \cite{MIK08,HHM10,M11,CON15,paper226}. For example, four-wave mixing produced by single-pass transmission through a single graphene layer has been observed \cite{HHM10}, while a second-order response at oblique incidence angles has been predicted \cite{M11}, and intrinsic second-order nonlinearities have been used to excite graphene plasmons from free-space beams via difference frequency generation \cite{CON15}. It has also been proposed that graphene nanostructures could enable quantum third-order nonlinearities \cite{paper226}.

We use a unified approach to determine the linear and nonlinear properties within the single-band approximation based upon the semi-classical Boltzmann transport equation \cite{MIK08,paper226,AM1976,M07_2}, which describes the evolution of the carrier distribution function $f_\mathbf{k}(\mathbf{r},t)$ at position $\mathbf{r}$ and momentum $\mathbf{k}$. Within this theory $\mathbf{k}$ and $\mathbf{r}$ obey the classical equations of motion: $\dot{\mathbf{r}} = \mathbf{v}_{\mathbf{k}} = (1/\hbar)\partial \epsilon_{\mathbf{k}}/\partial \mathbf{k}$, and $\hbar\dot{\mathbf{k}} = -e\mathbf{E}$. We are interested in energy scales $\lesssim 1$\,eV, so it is possible to linearize the graphene dispersion relation around the Dirac points,  $\epsilon_{\mathbf{k}} = \pm\hbar v_F|\mathbf{k}|$, where +(-) denotes doping to positive (negative) Fermi energies $E_F$. The single-band approximation holds provided that the optical frequency is less than $\sim$\,\,$2E_F$, such that absorption arising from interband electron-hole transitions is suppressed \cite{JBS09}. The carrier dynamics are then described by the equation
\begin{equation}
\label{eq:boltzmann_2}
\frac{\partial}{\partial t} f_\mathbf{k}(\mathbf{r},t) \pm v_F\hat{\mathbf{k}}\cdot\nabla_\mathbf{r}f_\mathbf{k}(\mathbf{r},t) = \frac{e}{\hbar}\mathbf{E}(\mathbf{r},t)\cdot\nabla_\mathbf{k} f_\mathbf{k}(\mathbf{r},t),
\end{equation}
where $\mathbf{E}$ is the sum of the external field $\mathbf{E}^\mathrm{ext}$ and the induced field $\mathbf{E}^\mathrm{ind}$ generated by the carrier distribution. 

The macroscopic quantities such as the density of charge and the surface current can be related to the microscopic dynamics of the carriers. For instance, the surface current depends on the microscopic carrier velocities as
\begin{equation}
\label{eq:current}
\mathbf{J}(\mathbf{r},t) = -e g_v g_s\int \frac{\mathrm{d}^2\mathbf{k}}{(2\pi)^2}\mathbf{v}_{\mathbf{k}}f_\mathbf{k}(\mathbf{r},t),
\end{equation}
where $g_s$\,=\,$g_v$\,=\,2 are the spin and valley degeneracies of graphene. The nonlinear set of equations (\ref{eq:boltzmann_2}) and (\ref{eq:current}) can be solved perturbatively to give the relation between the surface current and the electric field (i.e. the conductivity). At lowest order, one assumes that $f_\mathbf{k}$ is slightly displaced from its equilibrium (zero temperature) Fermi distribution, $f_\mathbf{k}^{(0)}(\mathbf{r},t) = \theta(k_F-k)$. Thus, one can substitute $f_\mathbf{k}^{(0)}$ into the term $\nabla_\mathbf{k} f_\mathbf{k}$ of equation (\ref{eq:boltzmann_2}), yielding a linear relationship between a perturbed distribution function $f^{(1)}$ and $\mathbf{E}$. Solving in the Fourier domain,  the perturbed distribution function $f^{(1)}$ can be inserted into equation (\ref{eq:current}) to find the resulting current. This yields a linear relation between the current and field, $\mathbf{J}(k,\omega)=\sigma^{(1)}(\omega)\,\mathbf{E}(k,\omega)$, where the proportionality constant is the familiar Drude conductivity \cite{HD07,WSS06}
\begin{equation}
\label{drude_1}
\sigma^{(1)}(\omega)=\frac{\ii e^2|E_F|}{\pi\hbar^2\omega}.
\end{equation} 
Finally, the procedure can be iterated by inserting the perturbed distribution function (e.g., $f^{(1)}$) into equation (\ref{eq:boltzmann_2}) to yield higher order conductivity functions, as derived in greater detail in \ref{AppNLC}.

Graphene is a centro-symmetric material, which is typically associated with a vanishing second-order nonlinearity \cite{boyd03}.\,\,Indeed, if the nonlinear response is spatially local, $J^{(2)}(2\omega,\mathbf{r}) = \sigma^{(2)}(\omega) E(\omega,\mathbf{r})^2$, spatial inversion symmetry implies that  $-J = \sigma^{(2)} (-E)^2$, which enforces that $\sigma^{(2)}=0$. This argument breaks down if the conductivity is nonlocal \cite{MICA11}, for example if $\sigma(\omega,q) \propto q$, such that the current depends on the electric field gradient, $J^{(2)} = \sigma^{(2)}(\omega) E\partial_\mathbf{r} E$.

In principle, nonlocal effects are present in any material. For a given electric field strength, the size of this nonlinear effect depends on a dimensionless parameter $k/k_{nl}$ \cite{heinz91}. Here $k$ is the wavevector of the light that dictates how rapidly the field changes in space, and $k_{nl}^{-1}$ is a characteristic length scale over which carriers in the material become sensitive to field gradients. In materials where the charges are tightly bound to their atoms, the relevant length scale $k_{nl}^{-1}$ is given by the atomic size of Angstroms, which is thus negligible compared to optical wavelengths. In conducting materials, the length scale is set by the typical distance between carriers, which is proportional to the inverse of the Fermi wavevector. In a typical metal like silver, the high carrier density also yields a negligible length scale of $k_{nl}^{-1} \sim k_F^{-1} \sim 1$ Angstrom. In contrast, in graphene we can simultaneously exploit two effects to increase significantly $k/k_{nl}$. First, graphene can be electrostatically tuned to have very low carrier densities to increase $k_{F}^{-1}$. Second, one can use tightly confined plasmon excitations in graphene, which have been shown to yield a reduction in the wavelength (or equivalently enhancement in wavevector $q_p$) compared to free-space light by two orders of magnitude. Indeed, below we show specifically that $k/k_{nl}\sim q_p/k_F\lesssim 1$ emerges as the relevant quantity to characterize the strength of nonlocal nonlinearities in graphene. 

After these considerations, we calculate the second-order conductivity using the procedure explained above. The second-order conductivity can be expanded in powers of $q_p$ in the long-wavelength limit, defined by the condition $v_F q_p/\omega_p \ll 1$ (see \ref{AppNLC}). As expected, the zeroth-order term, which corresponds to the local contribution, vanishes, while the term linear in $q_p$ provides a relation (in real space) between the electric fields at frequency $\omega_p$ and an induced current density at frequency $2\omega_p$
\begin{equation}
\label{eq:J}
J_i^{2\omega_p} = \sigma^{(1)}(2\omega_p)\,E_i^{2\omega_p} + \sigma_{ijkl}^{(2)}(2\omega_p;\omega_p)\,E_j^{\omega_p}\nabla_k E_l^{\omega_p}.
\end{equation}
Here $ijkl$ denote in-plane vector indices and summation over repeated indices is implied. The nonlocal second order conductivity tensor reads
\begin{equation}
\label{eq:J_2}
\sigma_{ijkl}^{(2)}(2\omega_p;\omega_p) = \mp\frac{\ii e^3 g_v g_s v^2_F}{32\pi\hbar^2\omega_p^3}\left(5\delta_{ij}\delta_{kl}-3\delta_{ik}\delta_{jl}+\delta_{il}\delta_{jk}\right).
\end{equation}
This result can be converted into a relation between the electrostatic potential and the induced charge, which reproduces previously obtained results for the nonlinear polarizability \cite{M11}.
\begin{figure}
\begin{center}
\includegraphics[width=125mm,angle=0,clip]{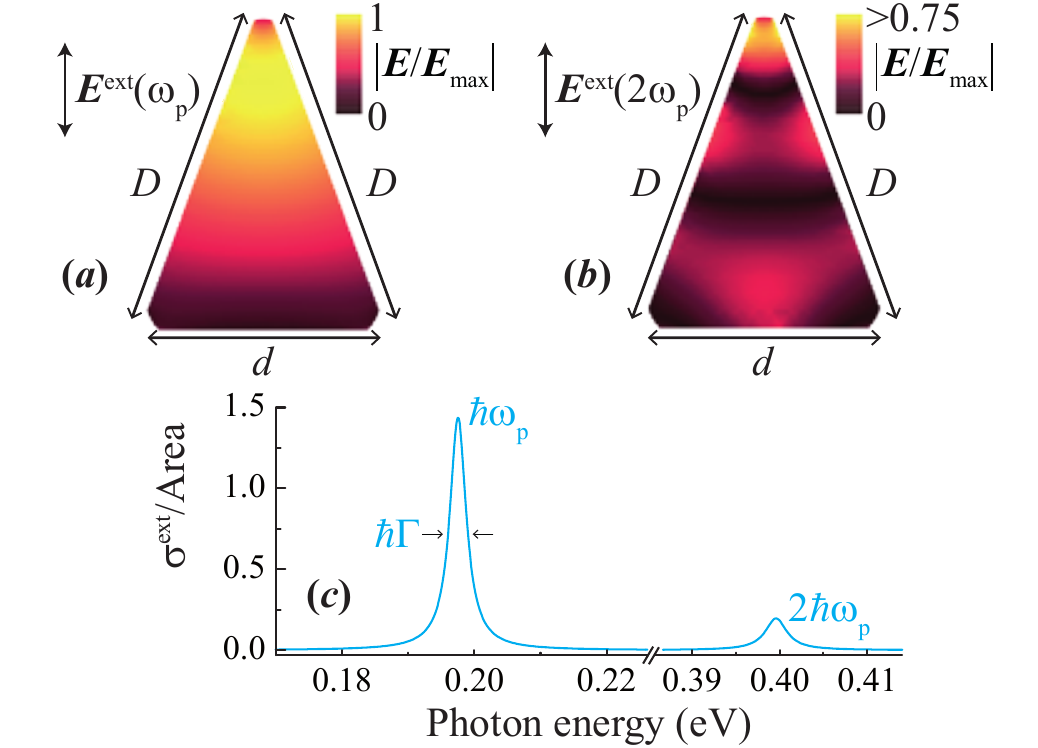}
\caption{{Plasmon modes in the graphene triangular nanoisland.} (a), (b) Induced electric field distribution associated with the first (a) and second (b) harmonic modes, respectively. The graphene structure consists of an isosceles triangle with side lengths $D=22$\,nm and $d=16.9$\,nm, and a doping level $E_F=0.2$\,eV with an intrinsic decay rate $\hbar\Gamma=3$\,meV (decay time $\sim$ 220 fs). (c) Extinction cross section normalized to the area ($S=169.6$\,$\mathrm{nm}^{2}$) of the triangles depicted in panels (a) and (b) with a strong fundamental dipolar mode and a secondary weaker dipolar mode.}
\label{fig_plasmons}
\end{center}
\end{figure}

\section{Quantum model of interacting graphene plasmons}

The Drude conductivity for infinite graphene given by equation (\ref{drude_1}) provides a valid description of the carrier dynamics of graphene when $\hbar\omega \lesssim E_F$ \cite{HD07,WSS06}, where the interband transitions can be neglected. Like any conductor in contact with a dielectric (or vacuum, as we assume here), graphene supports SPs with a dispersion relation given by
\begin{equation}
\label{disp_rel}
\frac{q_0}{q_p} \approx 2\alpha\,\frac{E_F}{\hbar\omega_p},
\end{equation}
where $q_0 = \omega/c$ is the free-space wavevector at the same frequency and $\alpha \approx 1/137$ is the fine structure constant. As $E_F\gtrsim \hbar\omega_p$, equation (\ref{disp_rel}) indicates a reduction in the plasmon  wavelength compared to free space by up to two orders of magnitude, which should significantly drive up the effects of spatially nonlocal interactions.

We have seen that at fixed field strength, the nonlinear interactions between plasmons in graphene should be increased due to a large ratio of $q_p/k_F$. However, what is most important for nonlinear optics is how to maximize the interaction strength per photon (i.e., per quantized plasmon). A simple argument, made more precise below, is that because the energy of a single plasmon is fixed at $\hbar\omega_p$, confining it to as small volumes $V$ as possible maximizes its intensity or electric field, $E_{0}\sim\sqrt{\hbar\omega_p/\epsilon_0 V}$. This motivates the study of nonlinear optical interactions between plasmons in nano-structures, which we now present in detail. As a specific example, we will focus on nanostructures that have plasmon resonances at frequencies $\omega_p$ and $2\omega_p$. This particular choice of structure is to facilitate DC or SHG.

The derivation of the quantum Hamiltonian of the system (reported in \ref{AppQUA}) starts from the expression of the electrostatic energy (a valid approach provided that the linear dimension of the structure $D$ is small compared to the free-space wavelength $\lambda_0$)
\begin{equation}
H = \frac{1}{2}\sum_{\omega_i}\int_{S} \mathrm{d}^2\mathbf{r}\,{\rho^{\omega_i}}^*(\mathbf{r})\phi^{\omega_i}(\mathbf{r}),
\end{equation}
where $\rho$ is the charge density and $\phi$ the electrostatic potential. The charge density can be replaced by the current density using the continuity equation, which in turn can be expressed in term of the electric field using equation (\ref{eq:J}). Expressing the potentials in terms of the electric fields we end up with an expression for the energy depending only on the electric field of the two modes, whose canonical quantization gives
\begin{equation}
\label{quantum_ham}
H = \hbar(\omega_p - \ii\Gamma_a'/2)\,a^\dagger a + \hbar(2\omega_p - \ii\Gamma_b'/2)\,b^\dagger b + \hbar g\left(\,b^\dagger a^2 + \mathrm{h.c.}\right).
\end{equation}
Here $a$ and $b$ are the annihilation operators of the two SP modes, and $g$ is an oscillation rate between a single plasmon with frequency $2\omega_p$ and two plasmons with frequency $\omega_p$ \cite{MAJ13}. Adopting a quantum jump approach we have added to the frequencies an imaginary part accounting for the total decay rates $\Gamma_a'$ and $\Gamma_b'$ of the two modes. The quantization associates with a single plasmon a typical electric field amplitude $E^{\omega_p}_0 \sim \left(\hbar\omega_p q_p  \big/ \epsilon_0 S\right)^{1/2}$, where $S$ is the structure area, confirming the large per-plasmon field associated with tight confinement.
The quantum coupling constant $g$ is rigorously given by the classical interaction energy between the nonlinear current at $2\omega_p$ and the fields at $\omega_p$, but with the classical field values replaced by the per-photon field strengths $\tilde{\mathbf{E}}^{\omega_i}(\mathbf{r})$
\begin{equation}
\label{eq:g_int}
\hbar g = \bigg|\frac{1}{4\ii\omega_p}\, \sigma_{ijkl}^{(2)}(2\omega_p;\omega_p)
\int_{S} \mathrm{d}^2\mathbf{r}\,\tilde{E}_i^{2\omega_p}(\mathbf{r})\tilde{E}_j^{\omega_p}(\mathbf{r})\nabla_k \tilde{E}_l^{\omega_p}(\mathbf{r})\bigg|.
\end{equation}
Equation (\ref{eq:g_int}) shows that $g$ is directly proportional to the second-order conductivity $\sigma_{ijkl}^{(2)}$ calculated in the previous section, and its dependence on the particular geometric configuration of the modes is confined to the overlap integral \cite{IRV06}. It should be noted that for extended graphene, the mode functions are simply propagating plane waves $E(\mathbf{r})\sim e^{\ii kz}$. Thus the integral in equation (\ref{eq:g_int}) produces a delta function, $g\propto \delta (2k_1-k_2)$, which reflects momentum conservation. In contrast, in small structures the spatially complex modes can be thought of as a superposition of many different wavevectors, and a large interaction strength is ensured by engineering the modes such that they have good spatial overlap \cite{RDG07}.

Using the fact that $E^{\omega_p}_0 \sim \left(\hbar\omega_p q_p  \big/ \epsilon_0 S\right)^{1/2}$, that the nonlinear conductivity has an amplitude $\sigma^{(2)}\sim e^3 v^2_F/\hbar^2\omega_p^3$, and that the field gradients occur over a length scale $q_p^{-1}$, one can readily verify that equation (\ref{eq:g_int}) predicts a general scaling of $g/\omega_p = \beta/(k_FD)^{7/4}$. The dimensionless coefficient of proportionality, which we call $\beta$, depends only on the geometric overlap of the modes (e.g., $\beta=0$ if the modes have the wrong symmetries, or $\beta\sim 1$ for modes with good overlap). As the minimum dimension of the structure should be comparable to the plasmon wavelength, $D\sim 1/q_p$, the maximum ratio of $g/\omega_p$ scales like $(q_p/k_F)^{7/4}$, confirming the enhanced nonlinearities as $q_p$ become comparable to $k_F$. Note that this relation is valid only for $q_p \lesssim k_F$, where the conductivity of graphene is Drude-like, as discussed above. In this derivation, we have assumed that a finite-size structure has the same conductivity as infinite graphene. Although this is not true for arbitrarily small structures, where quantum finite-size effects play a significant role, this approximation is already qualitatively correct for structures with $D\gtrsim 10$\,nm \cite{paper183}.

To show that a high overlap factor of $\beta\sim 1$ can be reached in typical structures, we consider one specific example of a doped graphene isosceles triangle embedded in vacuum. This choice enables a simple optimization to obtain the desired ratio of 2 between the SP mode frequencies. Indeed, we find that an aspect ratio $r=1.3$ produces plasmons at frequencies $\omega_p$ and $2\omega_p$ (see figure \ref{fig_plasmons}). The modes shown in figure\,\ref{fig_plasmons} are numerically computed using a commercial finite-difference code ($\rm{COMSOL}^{\textregistered}$) by driving the system with a plane wave whose associated external field $\mathbf{E}^{\mathrm{ext}}$ is polarized along the axis of symmetry of the triangle. We model the structure as a thin slab with rounded edges and a dielectric function $\epsilon=1+4\pi\ii\sigma^{(1)}/\omega t$. The thickness $t$ is chosen to be $t=0.5$\,nm~(this value is sufficiently small that the in-plane current has converged, and the results do not depend on the specific value), and the expression of $\sigma^{(1)}$ is given by the equation (\ref{drude_1}). Since the characteristic length of the structure is much smaller than the free-space wavelength, the response can be determined electrostatically, where the retardation and the response to the magnetic field are neglected. Furthermore, the ratio 1\,:\,2 between the first and second plasmon resonances is preserved independently of the actual size of the triangle and the doping \cite{paper235}. While the remaining parameters are somewhat arbitrary, as a numerical example, we consider the realistically achievable length and doping level of $D=22$\,nm and $E_F=0.2$\,eV. For this choice, we observe a pronounced first harmonic mode [figure\,\ref{fig_plasmons}(a),(c)] with energy $\hbar\omega_{p}\simeq0.20$\,eV, and a second harmonic resonance [figure\,\ref{fig_plasmons}(b),(c)] twice as energetic. Once we obtain the mode profiles, their nonlinear coupling is evaluated using the equations (\ref{eq:J_2}) and (\ref{eq:g_int}). Numerical calculations for this structure yield a value of $\beta=0.34$, hence the quantum oscillation rate $g$ reaches a remarkable 1.25\% value of the dipolar frequency $\omega_p$.

Surface plasmons in realistic graphene structures generally decay by non-radiative mechanisms, whose precise nature is still under active investigation \cite{paper212,NGM04,YLZ13}. We thus use a phenomenological description associating an intrinsic decay rate $\Gamma$ to the modes. For our numerical calculations we will assume a mode quality factor of $Q=\omega_p/\Gamma$ ranging from some tens to one hundred, close to what has been experimentally observed in nanostructures \cite{paper212}, although in our analytical results we will explicit keep track of the scaling with $\Gamma$.

In addition to intrinsic decay channels, graphene SPs can also be excited and detected through desirable channels, i.e., via radiative decay. We will use the notation $\kappa_{a,b}$ to indicate such decay rates. The total decay rate introduced in Hamiltonian (\ref{quantum_ham}) is thus $\Gamma_{a,b}' = \Gamma_{a,b} +  \kappa_{a,b}$. We will also introduce the notation $\eta_{a,b}$ to indicate the external coupling efficiencies of the modes, defined as $\kappa_{a,b}/\Gamma_{a,b}'$. For example, in our structure, the first and second harmonic modes radiate into free space at rates $\kappa_a \approx 2\times10^{-7}\omega_p$ and $\kappa_b \approx 5.4\times10^{-8}\omega_p$, as numerically calculated through the extinction cross sections of the incident field~(see \ref{AppDEC}). The external coupling efficiency can be increased by using more sophisticated techniques, such as SNOM \cite{paper196} or graphene nanoribbons \cite{paper226}.

\section{Observing and utilizing this nonlinearity: classical light}

The rate of oscillation or internal conversion between a single quantized plasmon and two lower-frequency plasmons is remarkable, particularly considering that the state-of-the-art down-conversion efficiency in conventional nonlinear crystals is $\sim 10^{-8}$ \cite{SAN11,GUE14}. It should be pointed out that the internal conversion rate holds independently of how the plasmons are generated. Of course, for both practical observation and for technological relevance, it would be ideal if the plasmons could be efficiently excited and subsequently converted back into propagating photons (such as from free space, fiber, or other evanescent modes). Motivated by this, we now examine the coupling problem to propagating photons in more detail and investigate how their intermediate conversion and interaction as plasmons manifests itself as strong, effective nonlinearities between propagating photons.
\begin{figure}
\begin{center}
\includegraphics[width=105mm,angle=0,clip]{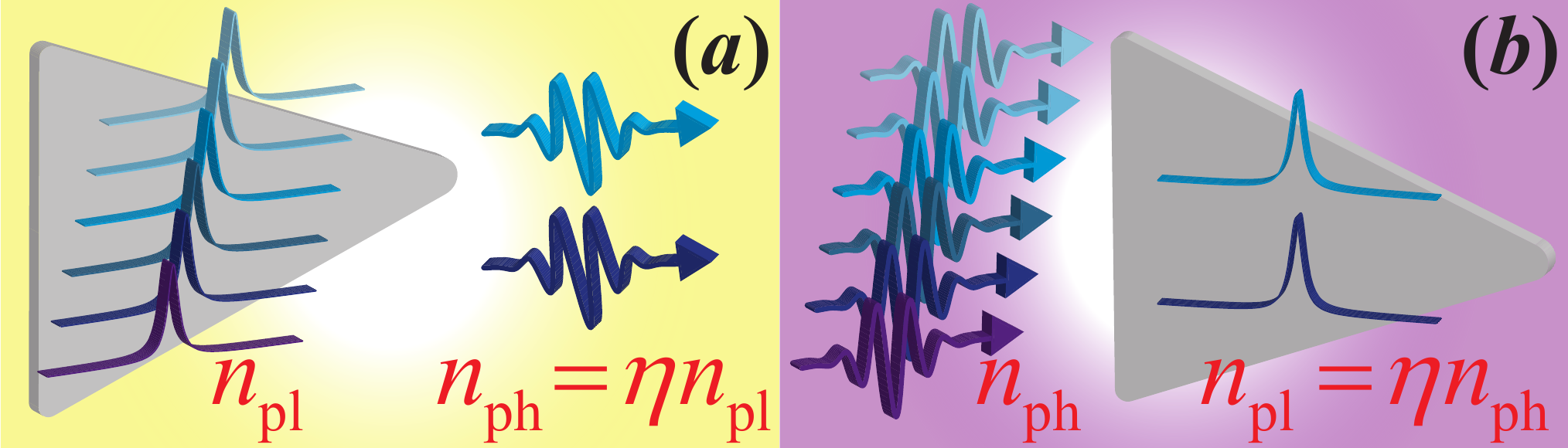}
\caption{(a) A given plasmon mode radiates into free space (or more generally, into any desirable channel) with an efficiency characterized by $\eta$. (b) By time-reversal symmetry, incoming photons in the same spatial mode excite plasmons with the same efficiency. The efficiency $\eta$ is related to the extinction cross section and free-space wavelength by $\eta = (2\pi/3)\sigma^{\mathrm{ext}}/\lambda_0^2$.}
\label{fig2}
\end{center}
\end{figure}

Remarkably, the extinction cross section $\sigma^{\mathrm{ext}} = (3/2\pi)\lambda_0^2\kappa/\Gamma'$ of a single nano-structure (see \ref{AppDEC}) can exceed its physical size. However, the low values of $\kappa/\Gamma'$ still imply that $\sigma^{ext}$ is much smaller than the diffraction limited area $\lambda^2_0$ for free-space beams, indicating that such sources cannot be used to excite plasmons efficiently. In particular, it can be shown using time-reversal symmetry that the best in-coupling (excitation) efficiency that can be achieved is the same as the out-coupling efficiency, $\eta$ \cite{GAF07}.
 The situation is illustrated schematically in figure~\ref{fig2}. This raises an important conceptual question. On one hand, graphene plasmons seem to represent the ``ultimate" quantum nonlinear optical device, capable of internal conversion at the single-photon level. However, very little incoming light enters the structure and turns into a plasmon, and vice versa, a small percentage of plasmons are radiated back into light. We now discuss various ways in which the strong quantum-level internal nonlinearities of graphene can be observed and utilized, given these limitations.

\begin{figure}
\begin{center}
\includegraphics[width=105mm,angle=0,clip]{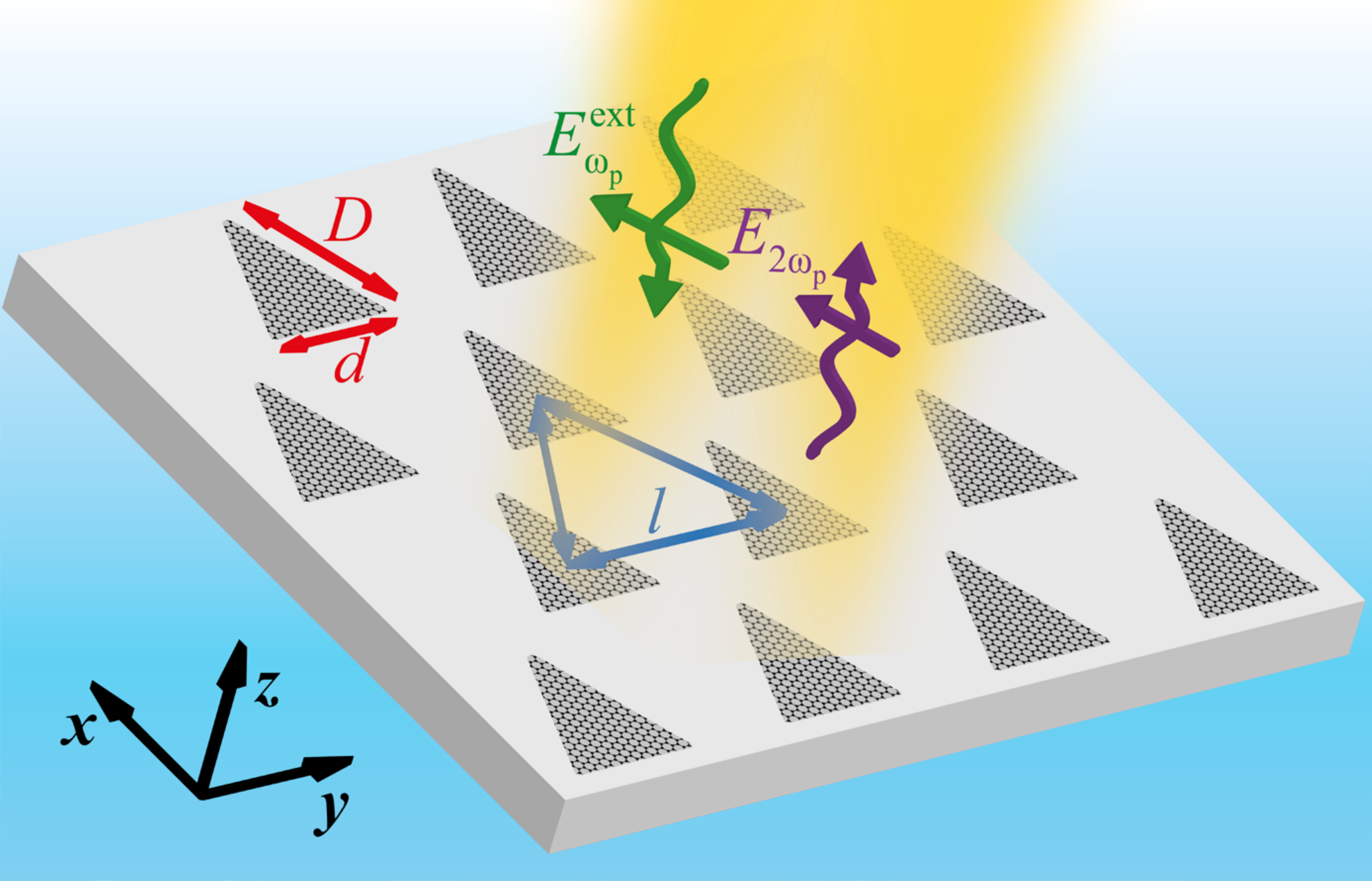}
\caption{A hexagonal array of triangular nanostructures illuminated by laser light at normal incidence and frequency $\omega_p$, resonant with the first plasmonic mode of the structures. The nonlinear coupling between this mode and the mode at frequency $2\omega_p$ generates an outgoing radiation field at this second harmonic, which is in a direction normal to the array.}
\label{fig3}
\end{center}
\end{figure}

One way of increasing the coupling to radiation, which has already been discussed in the linear optical regime, is to exploit an array of nano-structures \cite{paper212,paper182}. Intuitively, since the extinction cross section of a single element can exceed its physical size, having a dense array extending over an area larger than $\lambda^2_0$ guarantees efficient interaction with an incoming beam. We thus proceed to consider the nonlinear interaction between an incoming radiation field with frequency $\omega_p$ resonant with the fundamental mode and an array of nano-structures, as illustrated in figure~\ref{fig3}. We expect that the efficient coupling with an array will enable the incoming photons to excite plasmons at $\omega_p$, internally convert to plasmons at $2\omega_p$, and then re-radiate into free-space as a second harmonic signal. We consider here a hexagonal lattice of nanostructures with lattice period $l = 50$nm. The array is illuminated at normal incidence with a field of frequency $\omega$, and polarized along $\hat{x}$ to maximally drive the plasmon resonance (see figure~\ref{fig3}). 

From Hamiltonian (\ref{quantum_ham}) extended to include the coupling between the structures, we get the equations of motion of the operators for the first and second harmonic modes of structure $i$ in the array are 
\begin{eqnarray}
\label{array_eq}
\dot{a}_i =-\ii\left(\omega_p - \ii\Gamma_a'/2\right)a_i - \ii\frac{p_a}{\hbar}E^{\mathrm{ext}}_{\omega} - 2\ii g\,a_i^\dagger b_i + \ii\,\frac{p_a^2}{\hbar}\sum_j\,G^{\omega_p}_{ij}a_j,\\
\dot{b}_i =-\ii\left(2\omega_p - \ii\Gamma_b'/2\right)b_i - \ii\frac{p_b}{\hbar}E^{\mathrm{ext}}_{2\omega} - \ii g\,a_i^2 + \ii\,\frac{p_b^2}{\hbar}\sum_j\,G^{2\omega_p}_{ij}b_j,\nonumber
 \end{eqnarray} 
where the last term in both equations accounts for the dipole-dipole interaction with other nanostructures $j$ in the array. $G_{ij}=G(r_i,r_j)$ is the electromagnetic Green's function describing the field produced at position $r_i$ by a dimensionless dipole oscillating at $r_j$, while $p_a = \sqrt{3\pi\epsilon_0\hbar\kappa_ac^3/\omega_p^3}$ is the modulus of the electric dipole moment of a single plasmon in the first mode (an equivalent expression holds for $p_b$ at frequency $2\omega_p$). We have also included the possibility of driving either mode with classical free-space external fields, denoted by $E_{\omega}^{ext}$ and $E_{2\omega}^{ext}$.
\begin{figure}
\begin{center}
\includegraphics[width=105mm,angle=0,clip]{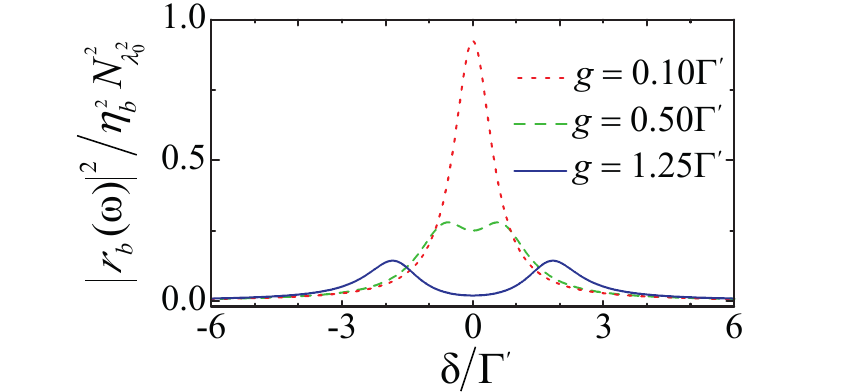}
\caption{{Back-scattered spectrum around $2\omega_p$.} Reflectance curves for a weak driving field as a function of the detuning $\delta$ (in units of the total decay rate $\Gamma'$) from the second mode of frequency $2\omega_p$, plotted for different values of the ratio $g/\Gamma'$. The value of the solid curve corresponds to the ratio $g/\Gamma' = 1.25$ that we have predicted theoretically for the structure presented in figure\,\ref{fig_plasmons}.}
\label{fig_trans}
\end{center}
\end{figure} 

Before considering the generation of a second harmonic, it is already interesting to point out that the strong internal interactions between plasmons can manifest itself in the \textit{linear} optical response to an incoming laser with frequency near the second mode $2\omega_p$. We proceed by solving the coupled system of equations (\ref{array_eq}) for a weak external driving field of frequency $\omega$ around $2\omega_p$. We consider specifically an approximation where edge effects are ignored (which becomes exact in the plane-wave limit and an infinite array), which makes the sum $\sum_j G_{ij}$ identical for each element. As discussed in detail in \ref{AppARR} the effect of the Green function is to renormalize both the resonance frequencies and the losses, so that $\omega_p \rightarrow \tilde{\omega}_p, \Gamma_a' \rightarrow \tilde{\Gamma}_a'$, etc. We find that the linear reflection coefficient of the array is
\begin{equation}
\label{r_b}
r_b(\omega) = -\frac{\ii\kappa_b N_{\lambda_0^2}}{2}\,\frac{\tilde{\delta}_a +\ii\tilde{\Gamma}_a'}{\left[\tilde{\delta}_a +\ii\tilde{\Gamma}_a'\right]\left[\tilde{\delta}_b + \ii\tilde{\Gamma}_b'/2\right] - 2g^2},
 \end{equation}
where $\tilde{\delta}_a = \omega - 2\tilde{\omega_p}$ is the detuning of the input field with respect to two times the renormalized first harmonic SP frequency, and similarly for $\tilde{\delta}_b$. The quantity $N_{\lambda_0^2} = (3/2\pi)(\lambda_0/2)^2/A$ is proportional to the number of structures in a diffraction limited area $\lambda_0^2$, as $A$ is the area of a unit cell in the array. In figure \ref{fig_trans}, we plot $|r_b(\omega)|^2$ as a function of the detuning for different values of the ratio $\Gamma'/g$. Here we have ignored the renormalized detunings, $\tilde{\delta}_i\rightarrow\delta_i$ for $i=a,b$, as the structure dimensions can be slightly altered to compensate for these shifts. We also take $Q=100$ and $Q=50$ for modes $a$ and $b$, respectively. Note that if the nonlinear interaction between plasmons is negligible ($g \ll \Gamma'$), the spectrum exhibits the typical Lorentzian peak associated with a resonant scatterer. We observe a qualitative difference in the reflection curve passing from the regime $g < \Gamma'/2$ to the regime in which $g > \Gamma'/2$, which is characterized by the appearance of a splitting in the reflection curve. Importantly, while an efficient external coupling increases the peak reflection of the structure, the magnitude of the mode splitting $2\sqrt{2}g$ does not depend on the coupling efficiency and represents a robust signature of quantum strong coupling between the SPs modes. We also emphasize that equation (\ref{r_b}) is only obtained by solving fully the equations (\ref{array_eq}), including quantum correlations between the two plasmon modes. Solving the classical limit, in which all quantum operators are replaced with numbers, would produce a Lorentzian spectrum for any value of $g$, which reinforces the appearance of a mode splitting as a quantum signature.

In a similar way, we can calculate the intensity emitted at frequency $2\omega_p$, when the system is driven at frequency $\omega_p$ by a classical external field. We find that the SHG signal intensity radiated into the far field is approximately (see \ref{AppARR} for the derivation)
\begin{eqnarray}
\label{intensity_m}
I_{2\omega_p}^{\rm{far}}&\approx&\frac{8g^2}{\hbar\omega_p{\Gamma_a'}^2\Gamma_b'}\frac{[\sigma^{\mathrm{ext}}_a]^2\sigma^{\mathrm{ext}}_b}{A^2}\,\left[I^{\mathrm{ext}}_{\omega_p}\right]^2, 
 \end{eqnarray} 
where $\sigma^{\mathrm{ext}}_{a,b}$ are the extinction cross sections of the two modes.
This expression is valid in the undepleted pump approximation, where the converted intensity is a small fraction of the incident. Using the previously quoted parameters for the triangular nanostructure, we find that a 1$\%$ conversion efficiency can be observed for the low driving intensity of roughly $10^{8}$ Wm$^{-2}$.

While we have presented here a semi-classical calculation, in which the input fields are treated as classical numbers, it would be interesting to find what is the conversion efficiency at the single-photon level. In particular, it would be interesting to see how graphene compares to the state-of-the-art efficiencies of $\sim 10^{-8}$ in bulk crystals for SHG of just a two-photon input. For this purpose, in the next section we use an approach based on the S-matrix formalism.

\section{Quantum frequency conversion}

In general, for a given few-photon input state, we wish to determine the effect of nonlinear interactions on the output. All of this information is contained in the S-matrix \cite{FAN10}, which specifically describes the overlap amplitude between a set of monochromatic incoming and outgoing freely propagating photons. Because monochromatic photons form a complete basis, the S-matrix thus contains all information about photon dynamics. In particular, it can be used to determine how a wave packet consisting of a superposition of monochromatic photons (i.e., decomposed into frequency components) interacts with the graphene nanostructure.

A simple example of an S-matrix element consists of the linear reflection amplitude $r_b(k)$ of a single photon of frequency $k_b$, which interacts with the higher-frequency SP mode (mode $b$), which we have calculated in the previous section by solving the Heisenberg equations of motion. In the S-matrix language the reflection coefficient corresponds to the matrix element between an incoming photon propagating in one direction (say to the right) and a photon of the same frequency $p_b=k_b$ scattered in the other direction (to the left). More compactly, this relation is formally written as $\bra{p^L_b} S\ket{k^R_b}\equiv r_b(k) \delta(k-p)$, where $\delta(k-p)$ denotes the Dirac delta function. Such an S-matrix element can be calculated by using standard input-output techniques~\cite{FAN10,GAD85}, which enable one to relate the outgoing field (after interaction) to the incoming field and internal dynamics of the nanostructure (governed by the Hamiltonian of equation (\ref{quantum_ham})). We assume that the incoming photon is focused at the diffraction limit, $S\sim\lambda_0^2$, and interacts with $N \equiv N_{\lambda_0^2}$ structures. In \ref{AppSCA}, we show that the resulting reflection coefficient gives a result of the form of equation (\ref{r_b}).

Analogously, we can express the amplitude for the DC process as the S-matrix element between an incoming photon of frequency $k_b$ near $2\omega_p$ and two outgoing photons of frequencies $p_a, q_a$ near $\omega_p$. For simplicity we study the case in which the incoming photon is a superposition of a photon coming from the right and one coming from the left so that we can avoid directional labels. We thus find for an array of $N$ structures
\begin{equation}
\label{eq:S_dc}
\bra{p_a,q_a} S\ket{k_b} = C\,\,r_b(k)\,r_a(p)\,r_a(q)\,\delta(k-p-q),
\end{equation}
where $r_a$, $r_b$ are respectively the reflection coefficients for photons in mode $a$ and $b$, and $C = 2Ng/\sqrt{2\pi\kappa_a^2\kappa_b}$~(see \ref{AppSCA}).

The S-matrix also enables one to calculate the dynamics of an incoming pulse. In particular, assuming a single-photon input wavepacket with a Fourier transform given by $f(k)$, we find that the total DC efficiency is given by $P_\mathrm{DC} = 1/2 \int \mathrm{d}p\,\mathrm{d}q\,|f(p+q)\,r_b(p+q)\,r_a(p)r_a(q)|^2$. For a near monochromatic resonant incoming photon, i.e., $|f(k)|^2 \approx \delta(k-2\omega_p)$, the result simplifies to
\begin{equation}
\label{eq:DC_prob_an}
P_\mathrm{DC} = \frac{16N^2\kappa_a^2\kappa_b\,g^2}{\Gamma_a'[\Gamma_a'\Gamma_b'+4g^2]^2},
\end{equation}
where $\Gamma' = \Gamma + N\kappa$.
The value of the coupling constant that maximizes the probability of conversion is $g = \sqrt{\Gamma_a'\Gamma_b'}/2$, for which we have
\begin{equation}
\label{eq:DC_prob_max}
P_\mathrm{DC} = N^2\,\bigg(\frac{\kappa_a}{\Gamma_a'}\bigg)^2\bigg(\frac{\kappa_b}{\Gamma_b'}\bigg).
\end{equation}

In general, we expect $g$ to exceed the plasmon linewidth in the graphene nanostructure considered, so that the condition $g = \sqrt{\Gamma_a'\Gamma_b'}/2$ is satisfiable, in contrast to conventional materials with weak nonlinear coefficients. For what concerns the optimal number of nanostructures, we identify two limits, one of low external coupling efficiency in which the array-enhanced external coupling does not overcome the losses i.e., $\kappa \ll \Gamma$, and the opposite case in which $\kappa \gtrsim \Gamma$. In the first limit, which is satisfied for the system parameters presented earlier, the total decay rate $\Gamma'$ is roughly independent of the number of structures $N$ and $P^\mathrm{max}_\mathrm{DC} \approx N^2\eta_a^2\,\eta_b$ (we recall again that $\eta_i = \kappa_i/\Gamma_i$). It is clear that in this limit the use of an array of nanostructures is an efficient way to increase the conversion (which anyway remains much smaller than 1). For our system parameters, we find that $P_{DC}^\mathrm{max}\approx 10^{-7}$, which compares favorably with state-of-the-art numbers $\sim10^{-8}$, a surprising result considering that graphene is not a bulk nonlinear crystal. In the opposite limit of good external coupling we find that $P^\mathrm{max}_\mathrm{DC} = N^{-1}\eta_a^2\,\eta_b$. This remarkable result indicates that ultimately, there is a fundamental inequivalence between using many structures to increase the (linear) response, and working to improve the coupling to just a single structure. In particular, in the limit of efficient coupling, the strong nonlinear interaction between plasmons becomes diluted by having multiple structures. Intuitively, this $N^{-1}$ scaling can be understood from the complementary process of SHG (whose S-matrix is identical to DC, as shown later). Clearly, in order for two incoming photons to create a second harmonic, they must excite two plasmons in the same structure. However, with many structures, the probability that this occurs (i.e., compared to exciting single plasmons in two different structures) falls like $N^{-1}$. We thus argue that the development of techniques \cite{paper196,paper226} to efficiently couple to single structures is of fundamental importance to take maximal advantage of the strong intrinsic nonlinear interactions between graphene plasmons.

It should further be noted that the created photon pairs are frequency-entangled [see equation\,(\ref{eq:S_dc})], as energy conservation requires that the sum of their frequencies equals that of the incoming single photon.
\begin{figure}
\begin{center}
\includegraphics[width=125mm,angle=0,clip]{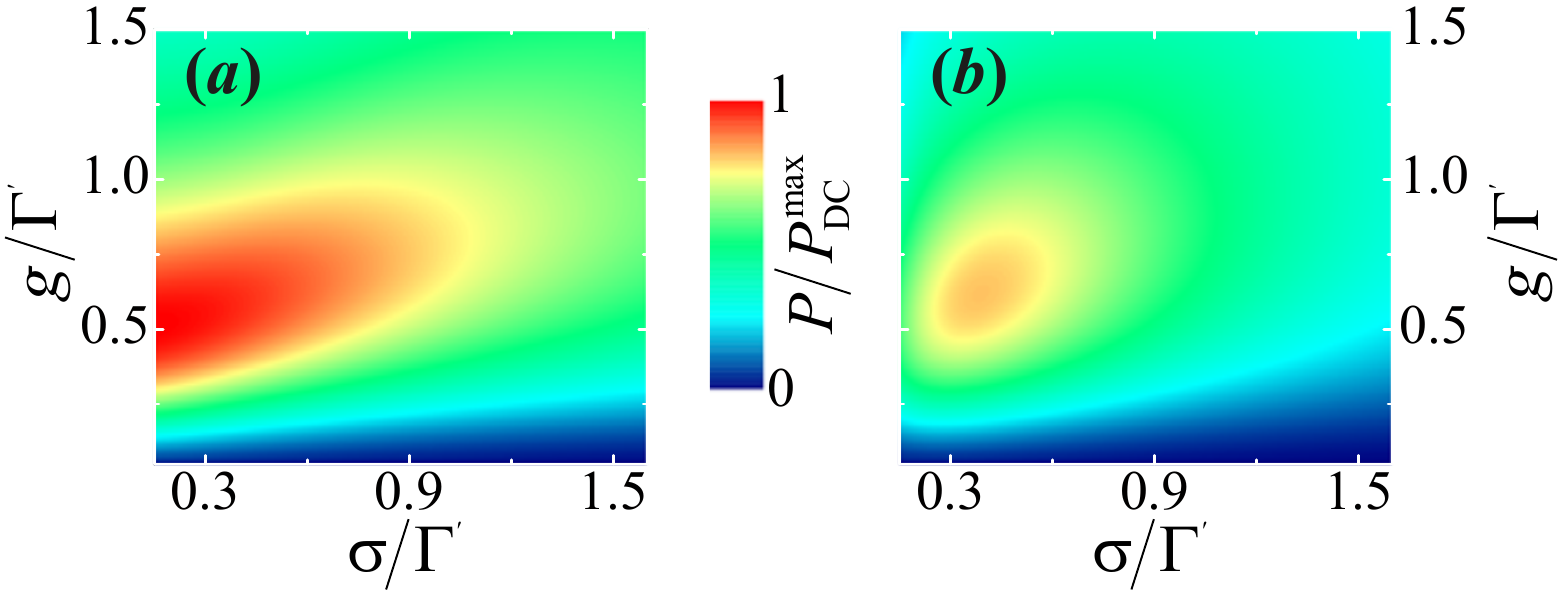}
\caption{{ Single photon DC and SHG normalized probabilities.} (a) Probability of DC for a photon in a Gaussian wavepacket of center frequency $2\omega_p$ and bandwidth $\sigma$. $P_\mathrm{DC}$ is plotted as function of $\sigma$ and $g$ (in units of $\Gamma'$), and normalized with respect to $P^\mathrm{max}_\mathrm{DC}$. (b) Probability of SHG for a pair of uncorrelated photons in Gaussian wavepackets of center frequency $\omega_p$ and bandwidth $\sigma$, normalized as in (a).}
\label{fig_efficiencies}
\end{center}
\end{figure}
Intuitively, one expects that the DC process remains efficient as long as the incoming pulse bandwidth $\sigma$ is smaller than the cavity linewidth $\Gamma'$. This can be seen quantitatively in figure \ref{fig_efficiencies}(a), where Gaussian single-photon inputs with bandwidth $\sigma$ are considered, i.e., $f(k) \propto e^{-(k-\omega_p)^2/4\sigma^2}$.

In SHG  two photons with frequencies centered around $\omega_p$ are (partially) converted in a single photon of frequency $2\omega_p$. By the time reversal symmetry of the scattering matrix the relation $\bra{p_a,q_a} S\ket{k_b} = \bra{k_b} S\ket{p_a,q_a}^*$ holds. This implies that in principle, a maximum up-conversion efficiency of $P_\mathrm{SHG}^{\mathrm{max}} = P_\mathrm{DC}^{\mathrm{max}}$ can be achieved, but only if the two-photon input itself is an entangled state. In figure \ref{fig_efficiencies}(b), we consider the more realistic case of two identical, separate photons, each represented as a Gaussian pulse of width $\sigma$. It can be noticed the qualitatively different functional behavior of $P_\mathrm{DC}$ and $P_\mathrm{SHG}$. The latter saturates at a lower value than the former and exhibits a maximum for a finite value of $\sigma$, going to zero for both the limits $\sigma \rightarrow 0$ and $\sigma \rightarrow \infty$. The inability to deterministically up-convert two separate photons ($P_\mathrm{SHG}=1$), even for perfect coupling efficiencies, notably deviates from the semiclassical prediction that perfect conversion can be achieved \cite{RDG07}.

\begin{figure}
\begin{center}
\includegraphics[width=140mm,angle=0,clip]{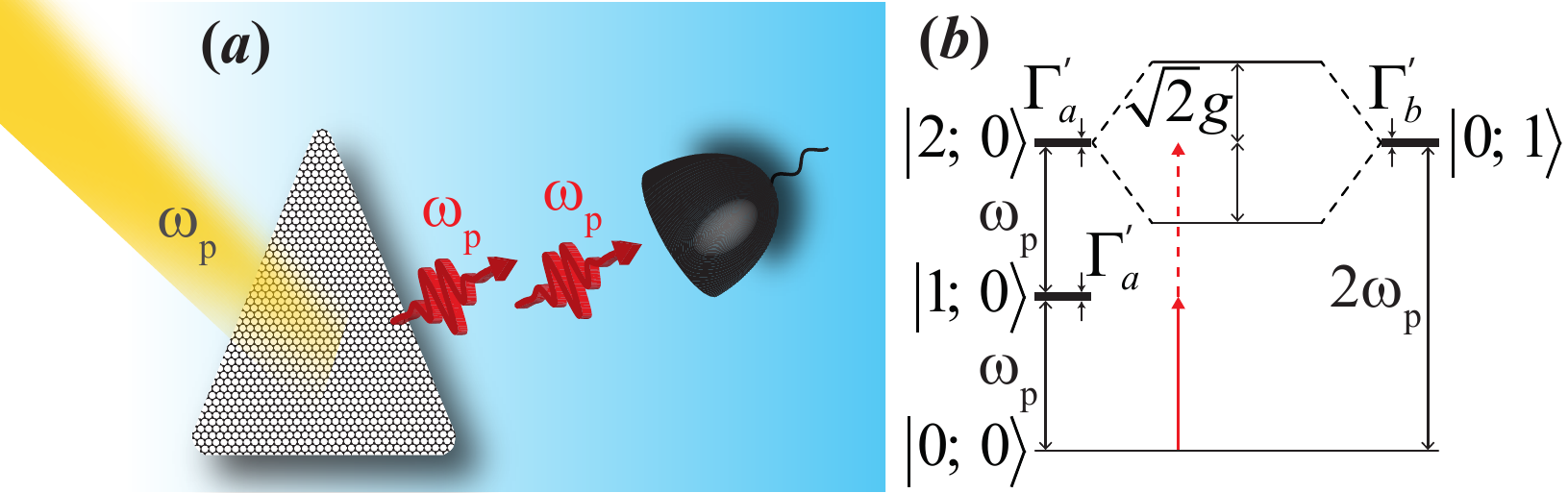}
\caption{{Quantum light generation in a graphene triangular nanoisland.} (a) Schematic showing the creation of non-classical light. A coherent state beam~(yellow) of frequency $\omega_p$ incident on the graphene is scattered and produces anti-bunched light~(red). (b) Energy level structure of the system, where the notation $\ket{m,n}$ indicates the occupation of $m\,(n)$ plasmons in mode $\omega_p\,(2\omega_p)$. The dressed states generated by the coupling between $\ket{2;0}$ and $\ket{0;1}$ are also represented. Red arrows illustrate the origin of photon blockade. Due to the nonlinear coupling, the nominally degenerate states $\ket{2;0}$ and $\ket{0;1}$ hybridize into two dressed states with frequencies $2\omega_p \pm g/\sqrt{2}$. When the fundamental mode is resonantly driven, the population of that mode by a single photon (solid red arrow) blocks the excitation of a second photon (dashed red arrow), as the mode hybridization results in the absence of a state at $2\omega_p$.
}
\label{fig_scheme}
\end{center}
\end{figure}
We conclude showing that a single graphene nanostructure can generate nonclassical light when irradiated with weak classical light at the lower frequency. We have seen above that in the strong quantum coupling regime, $g > \Gamma'/2$, a mode splitting at the second resonance appears. Physically, this splitting arises because the nonlinear interaction given in the Hamiltonian of equation (\ref{quantum_ham}) strongly mixes a single photon $\ket{0;1}$ in mode $2\omega_p$ with two photons $\ket{2;0}$ in mode $\omega_p$, as shown in figure~\ref{fig_scheme}(b). The resulting eigenstates of the Hamiltonian are symmetric and antisymmetric combinations $\ket{0;1}\pm\ket{2;0}$ with frequencies $2\omega_p \pm \sqrt{2}g$.
The mode splitting creates an effective nonlinearity: once a single plasmon of frequency $\omega_p$ enters the system, the absence of a resonant state at $2\omega_p$ prevents a second plasmon from entering, creating a blockade effect \cite{MAJ13}. This is a complementary signature of strong coupling observable in the lower mode. It can be quantified by considering the second-order correlation function of back-scattered photons (for instance left-propagating photons when the system is driven by right-propagating laser light) $g^{(2)}(t) = \braket{a_{L,\mathrm{out}}^\dagger(\tau)a_{L,\mathrm{out}}^\dagger(\tau+t)a_{L,\mathrm{out}}(\tau+t)a_{L,\mathrm{out}}(\tau)}/\braket{a_{L,\mathrm{out}}^\dagger(\tau)a_{L,\mathrm{out}}(\tau)}^2$. The output field itself is related to the input field and plasmon mode by the equation $a_{L,\mathrm{out}} = a_{L,\mathrm{in}} + \sqrt{\kappa_a/2}\,a$. However, as the left-going input field is in the vacuum state, the corresponding input operator has no effect. Thus the second-order correlation function can be written directly in terms of the plasmon mode $a$, $g^{(2)}(t) = \braket{a^{\dagger}(\tau)a^{\dagger}(\tau+t)a(\tau+t)a(\tau)}/\braket{a^{\dagger}(\tau)a(\tau)}^2$. For $t = 0$ this function indicates the relative probability to detect two photons at the same time. Values of $g^{(2)}(0) < 1$ indicate the presence of nonclassical light. In the limit of weak driving amplitude we find that
\begin{equation}
g_{a}^{(2)}(0) = \frac{{\Gamma'}^2(16g^2+3{\Gamma'}^2)}{3(4g^2+{\Gamma'}^2)^2}.
\end{equation}
For $g=0$ it acquires a value of $g_a^{(2)}(0)=1$, reflecting the coherent state statistics of the laser, while exhibiting strong anti-bunching $(g_a^{(2)}(0)<1)$ when $g \gtrsim \Gamma'/2$.
It is particularly important that $g_{a}^{(2)}(0)$ is independent of the external coupling efficiency $\kappa/\Gamma'$, thus making this effect a robust signature of strong quantum coupling between plasmon modes.

\section{Outlook and conclusion}

We have shown that second-order nonlinear optical interactions between plasmons in graphene nanostructures can be remarkably strong. Signatures of such nonlinearities should be immediately observable in experiments involving arrays of nanostructures, where incident free-space light can undergo frequency mixing at very low input powers via interaction with plasmons.

 We further show that single nanostructures should exhibit the capability to generate non-classical states of light, observable even with low coupling efficiencies, which opens up a novel route to quantum optics as compared to the conventional approach of using atom-like emitters. With improved coupling efficiencies to the modes of these nanostructures, it would become possible to realize efficient second-harmonic generation or down-conversion at the level of a few quanta, which would exceed the capabilities of current systems by several orders of magnitude. While we focused on one concrete example consisting of a graphene nanotriangle, our conclusions are quite adaptable. Thus, it would be interesting to explore further the potential of this unique ``nonlinear crystal" in a wide variety of classical and quantum nonlinear optical devices. It would also be interesting to investigate the nonlinear optical response of even smaller structures \cite{paper247,COX15}, which is expected to deviate significantly from large-scale graphene due to quantum finite-size effects. Finally, we anticipate that our work will open up the intriguing possibility of a search for new materials that are capable of attaining the quantum nonlinear regime.

\ack

The authors thank M. Jablan, J. D. Cox and F. H. L. Koppens for useful discussions. DEC acknowledges support from Fundacio Privada Cellex Barcelona. MTM acknowledges the international PhD-fellowship program ``la Caixa-Severo Ochoa". All authors acknowledge support from European Project GRASP.

\appendix

\section{Semiclassical derivation of the second-order conductivity}
\label{AppNLC}

We calculate in this section of the Appendix the second-order conductivity of extended graphene. In the main text, we apply this result to a finite-size structure. In principle, it should be noted that the response of a finite-size structure is not spatially homogeneous, and applying the conductivity functions for extended graphene is an approximation which likely gives an error of $\sim \lambda_F/\lambda_p$. An exact treatment for finite structures requires the calculation of the response function, which involves intensive numerical computations \cite{paper247}.

As explained in the main text, we use a semiclassical single-band approach describing the dynamics of carriers in graphene, which is nominally a zero-gap semiconductor. However, by techniques such as electrostatic gating \cite{CGP09}, one band can become partly filled with carriers. We characterize the carriers within this band by the distribution function $f_\mathbf{k}(\mathbf{r},t)$ which is defined so that
\begin{equation}
\mathrm{d}N = f_\mathbf{k}(\mathbf{r},t)\,\mathrm{d}^2\mathbf{k}\,\mathrm{d}^2\mathbf{r}
\end{equation}
is the number of carriers with positions lying within a surface element $\mathrm{d}^2\mathbf{r}$ about $\mathbf{r}$ and momenta lying within a momentum space element $\mathrm{d}^2\mathbf{k}$ about $\mathbf{k}$, at time $t$. When collisions between carriers are neglected, a conservation equation is satisfied by the function $f_\mathbf{k}(\mathbf{r},t)$ (equation\,(\ref{eq:boltzmann_2}) of the main text \cite{AM1976}). In Fourier space it can be written in the form
\begin{equation}
\label{eq:boltzmann_3}
f_\mathbf{k}(\mathbf{q},\omega) = \frac{\mathrm{i}e}{\hbar(\omega \mp v_F\hat{\mathbf{k}}\cdot\mathbf{q})}
\int \frac{\rm{d}^2\mathbf{p}}{(2\pi)^2}\int^{\infty}_{-\infty}\frac{\mathrm{d}\nu}{2\pi}\mathbf{E}(\mathbf{q}-\mathbf{p},\omega-\nu)\cdot\frac{\partial f_\mathbf{k}(\mathbf{p},\nu)}{\partial \mathbf{k}},
\end{equation}
which exhibits a nonlinear character. We assume that the electric field perturbs the equilibrium distribution weakly so that we can solve the equation\,(\ref{eq:boltzmann_2}) iteratively, obtaining a perturbation series in $\mathbf{E}$ for the distribution function.

At zero order, ignoring finite-size effects as noted above, we simply replace the distribution function on the RHS with the (zero temperature) Fermi distribution $f_\mathbf{k}^{(0)}(\mathbf{r},t) = \theta(k_F-k)$, obtaining as solution the first order contribution to the conductivity, which is linear in the electric field:
\begin{equation}
\label{eq:f_1}
f^{(1)}_\mathbf{k}(\mathbf{q},\omega) = -\frac{\mathrm{i}e\hat{\mathbf{k}}\cdot\mathbf{E}(\mathbf{q},\omega) }{\hbar(\omega \mp v_F \hat{\mathbf{k}}\cdot\mathbf{q})}\delta(k - k_F).
\end{equation}
In turn, inserting equation\,(\ref{eq:f_1}) into equation\,(\ref{eq:boltzmann_2}) of the main text, we get the second-order contribution
\begin{eqnarray}
\label{eq:f_2}
\fl f^{(2)}_\mathbf{k}(\mathbf{q},\omega)=& \frac{e^2}{\hbar(\omega \mp v_F \hat{\mathbf{k}}\cdot\mathbf{q})}\,\times\\
 &\int \frac{\mathrm{d}^2\mathbf{p}}{(2\pi)^2}\int^{\infty}_{-\infty}\frac{\mathrm{d}\nu}{2\pi}\,\mathbf{E}(\mathbf{q}-\mathbf{p},\omega-\nu)\cdot\frac{\partial }{\partial \mathbf{k}}\left\{\frac{\hat{\mathbf{k}}\cdot\mathbf{E}(\mathbf{p},\nu)}{\hbar(\nu \mp v_F \hat{\mathbf{k}}\cdot\mathbf{p})}\delta(k - k_F)\right\} \nonumber
\end{eqnarray}
Moreover equation\,(\ref{eq:current}) of the main text
\begin{equation}
\mathbf{J}(\mathbf{r},t) = -e g_v g_s\int \frac{\mathrm{d}^2\mathbf{k}}{(2\pi)^2}\mathbf{v}_{\mathbf{k}}f_\mathbf{k}(\mathbf{r},t),
\end{equation}
provides a relation between the macroscopic density of electric current and the microscopic distribution function. Inserting the elements of the series we got for the distribution function, and taking into account the definition of $n^{th}$-order conductivity, we obtain that
\begin{equation}
\label{drude}
\sigma^{(1)}_{ij}(\mathbf{q},\omega)  =  \frac{\mathrm{i}e^2 g_v g_s v_F}{\hbar}\int \frac{\mathrm{d}^2\mathbf{k}}{(2\pi)^2}\frac{k_ik_j}{k^2(\omega \mp v_F\hat{\mathbf{k}}\cdot\mathbf{q})}\delta(k - k_F),
\end{equation}
and
\begin{eqnarray}
\label{eq:sigma_2}
\fl \sigma^{(2)}_{ilm}(\mathbf{q},\omega;\mathbf{q}-\mathbf{p},\omega-\nu,\mathbf{p},\nu) = \frac{e^3 g_v g_s v_F}{\hbar^2}\int \frac{\mathrm{d}^2\mathbf{k}}{(2\pi)^2}\,\frac{1}{k^2(\omega \mp v_F \hat{\mathbf{k}}\cdot\mathbf{q})(\nu \mp v_F \hat{\mathbf{k}}\cdot\mathbf{p})}\times 
 \\ 
  \left[\delta_{il}k_m - \frac{\omega}{k^2(\omega \mp v_F \hat{\mathbf{k}}\cdot\mathbf{q})}k_ik_lk_m \pm \frac{v_F}{k(\omega \mp v_F \hat{\mathbf{k}}\cdot\mathbf{q})}k_iq_lk_m\right]\delta(k - k_F),\nonumber  
\end{eqnarray}
where the integration over $\mathbf{k}$ is on a circle of radius $k_F$, because of the linearization of the band.

Analytical results can be obtained in the long-wavelength limit ($v_F q/\omega \ll 1$), by expanding the denominators in $q$ and $p$. The dominant contribution to the linear conductivity is the zero-order one, giving rise to the well-known local Drude conductivity of graphene displayed in the equation (\ref{drude_1}) of the main text.

For the second-order conductivity, it can be easily proven that the zero-order expansion in $q$ and $p$ of equation\,(\ref{eq:sigma_2}) gives a vanishing contribution when the integral is performed, so that the dominant term is the first-order expansion in $q$ and $p$, which corresponds to a nonlocal contribution. 

The results are given by
\begin{eqnarray}
 \sigma^{(2)}_{ilm}(\mathbf{q},\omega;\mathbf{q}-\mathbf{p},\omega-\nu,\mathbf{p},\nu) =&\pm\frac{e^3 g_v g_s v^2_F}{16\pi\hbar^2}\bigg[\frac{2}{\omega^2\nu}\big(\delta_{il}q_m - \delta_{lm}q_i + \delta_{im}q_l) \nonumber \\
 & +\frac{1}{\omega\nu^2}\big(3\delta_{il}p_m - \delta_{lm}p_i -\delta_{im}p_l)\bigg].
\end{eqnarray}
With the formula for the conductivity above, the expression
\begin{eqnarray}
\label{eq:J_4}
J_i^{(2)}(\mathbf{q},\omega) =& \int \frac{\rm{d}^2\mathbf{p}}{(2\pi)^2}\int^{\infty}_{-\infty}\frac{\mathrm{d}\nu}{2\pi}\,\sigma^{(2)}_{ilm}(\mathbf{q},\omega;\mathbf{q}-\mathbf{p},\omega-\nu,\mathbf{p},\nu)\,\times  \\ 
& E_l(\mathbf{q}-\mathbf{p},\omega-\nu)\,E_m(\mathbf{p},\nu),\nonumber
\end{eqnarray}
can be Fourier transformed back to the real space, resulting finally in the equations (\ref{eq:J}) and (\ref{eq:J_2}) of the main text.

\section{Quantization of the two-mode structure energy and derivation of the conversion rate $g$}
\label{AppQUA}

In the limit $D/\lambda_0 \ll 1$, where $D$ is the linear dimension of the graphene structure and $\lambda_0$ is the incident radiation wavelength, the electrostatic approximation can be assumed. Thus, the total energy present in the structure is given by
\begin{equation}
H = \frac{1}{2}\sum_{\omega_i}\int_{S} \mathrm{d}^2\mathbf{r}\,{\rho^{\omega_i}}^*(\mathbf{r})\phi^{\omega_i}(\mathbf{r}),
\end{equation}
where $\rho$ is the charge density and $\phi$ the electrostatic potential. Implementing the continuity equation, the relation between the potential and the electric field, and writing explicitly the two frequencies contributions, we get that the total energy is given by
\begin{equation}
\label{eq:clas_ham}
H = \frac{1}{2\ii\omega_p}\int_{S} \mathrm{d}^2\mathbf{r}\,{J^{\omega_p}_i}^*(\mathbf{r})E^{\omega_p}_i(\mathbf{r})  + \frac{1}{4\ii\omega_p}\int_{S} \mathrm{d}^2\mathbf{r}\,{J^{2\omega_p}_i}^*(\mathbf{r})E^{2\omega_p}_i(\mathbf{r}).
\end{equation}
The currents can be expressed in terms of the electric fields: $J_i^{\omega_p} = \sigma^{(1)}(\omega_p)E_i^{\omega_p}$, while $J_i^{2\omega_p}$ is given by equation (\ref{eq:J}) of the main text.

At this point, we can impose the quantization condition to the first mode:
\begin{equation}
\frac{\sigma^{(1)}(\omega_p)}{2\ii\omega_p}\int_{S} \mathrm{d}^2\mathbf{r}\,|E^{\omega_p}_i(\mathbf{r})|^2 = \hbar\omega_p\,a^\dagger a,
\end{equation}
which can be enforced with the substitution $E_i^{\omega_p}(\mathbf{r}) \rightarrow \tilde{E}_i^{\omega_p}(\mathbf{r})\,a = E^{\omega_p}_0\,f_i^{\omega_p}(\mathbf{r})\,a$. Here, $f^{\omega_p}(\mathbf{r})$ is a vectorial function which describes the geometry of the mode and normalized such that $\max |f^{\omega_p}(\mathbf{r})| = 1$, $E^{\omega_p}_0 = \left(\hbar\omega_p q_p  \big/ \epsilon_0 S\mu\right)^{1/2}$ is the maximum single-photon electric field amplitude, and $\mu =  S_\mathrm{eff}^{\omega_p}/S$,  with $S_\mathrm{eff}^{\omega_p} = \int_{S} \mathrm{d}^2\mathbf{r}\,|f^{\omega_p}_i(\mathbf{r})|^2$ is the ratio between the effective mode area and the physical area of the structure.

Now we consider the mode at frequency $2\omega_p$. We substitute the current shown in equation (\ref{eq:J}) into equation (\ref{eq:clas_ham}), and similarly define a single-photon electric field amplitude and annihilation operator $b$ for this mode. This procedure yields a non-interacting term in the Hamiltonian, $2\hbar\omega_p b^{\dagger}b$, and an interacting term
\begin{equation}
\label{eq:g_sm}
 H_\mathrm{int} = \frac{1}{4\ii\omega_p}\,\sigma_{ijkl}^{(2)}(2\omega_p;\omega_p)\int_{S} \mathrm{d}^2\mathbf{r}\,[{\tilde{E}_i^{2\omega_p}}(\mathbf{r})]^*\tilde{E}_j^{\omega_p}(\mathbf{r})\nabla_k \tilde{E}_l^{\omega_p}(\mathbf{r})\,a^2b^\dagger + \mathrm{h.c.}
\end{equation}
Note that using the freedom in the definition of the phases of $\tilde{E}_i^{\omega_p}$ and $\tilde{E}_i^{2\omega_p}$ we can make the expression in front of $a^2b^\dagger$, $\hbar g$, to be real, obtaining in this way equation (\ref{eq:g_int}) of the main text.

If we express $\tilde{E}_l^{\omega_p}(\mathbf{r})$ as $E^{\omega_p}_0\,f_i^{\omega_p}(\mathbf{r})$, in order to separate the geometric part of the problem, we get that the ratio between $g$ and $\omega_p$ is equal to
\begin{eqnarray}
\label{eq:g_sm_2}
\fl
\frac{g}{\omega_p}&=& \frac{1}{4\ii\hbar\omega_p^2}\,\sigma_{ijkl}^{(2)}(2\omega_p;\omega_p)\,\left[E^{\omega_p}_0\right]^2\,E^{2\omega_p}_0\int_{S} \mathrm{d}^2\mathbf{r}\,{f_i^{2\omega_p}}^*(\mathbf{r})f_j^{\omega_p}(\mathbf{r})\nabla_k f_l^{\omega_p}(\mathbf{r})  \\
\fl
 &=&\mp\frac{1}{4\ii\hbar\omega_p^2}\,\frac{\ii e^3v^2_F}{8\pi\hbar^2\omega_p^3}\,\frac{2\pi\hbar^3\omega_p^3}{e^2|E_F|S_\mathrm{eff}^{\omega_p}}\left(\frac{16\pi\hbar^3\omega_p^3}{e^2|E_F|S_\mathrm{eff}^{2\omega_p}}\right)^{1/2}\,\Theta_{ijkl}\int_{S} \mathrm{d}^2\mathbf{r}\,{f_i^{2\omega_p}}^*(\mathbf{r})f_j^{\omega_p}(\mathbf{r})\nabla_k f_l^{\omega_p}(\mathbf{r}),\nonumber
\end{eqnarray}
where we have called the tensorial part of the second-order conductivity $\Theta_{ijkl} = \left(5\delta_{ij}\delta_{kl}-3\delta_{ik}\delta_{jl}+\delta_{il}\delta_{jk}\right)$. Now we make the lengths in the integral dimensionless, introducing $\mathbf{\hat{r}} = \mathbf{r}/D$ and $\hat{\nabla} = D\nabla$. Furthermore, we introduce the dimensionless parameter $\xi_1 = D^3/S_\mathrm{eff}^{\omega_p}[S_\mathrm{eff}^{2\omega_p}]^{1/2}$. Implementing the former in equation (\ref{eq:g_sm_2}), we obtain the final expression
\begin{eqnarray}
\label{eq:g_sm_3}
\fl
 \frac{g}{\omega_p}=\mp\frac{v_F^2\hbar^2}{4|E_F| D^2}\left(\frac{\pi}{\hbar\omega_p|E_F|}\right)^{1/2}\,\xi_1\,\Theta_{ijkl}\int_{S} \mathrm{d}^2\mathbf{\hat{r}}\,f_i^{2\omega_p}(\mathbf{\hat{r}})f_j^{\omega_p}(\mathbf{\hat{r}})\hat{\nabla}_k f_l^{\omega_p}(\mathbf{\tilde{r}}),
\end{eqnarray}
Note that now the integral in the last expression is fully dimensionless and depends only on the geometry of the two modes.

In a general way, the plasmon frequency $\omega_p$ is related to the dimension of the structure and the doping by the relation \cite{paper235}
 \begin{equation}
\omega_p = \xi_2\left(\frac{2\alpha c}{\hbar}\frac{E_F}{D}\right)^{1/2},
\end{equation}
where $\xi_2$ is a factor of order one that depends only on the shape of the structure considered, $\alpha$ is the fine-structure constant, and $c$ is the speed of light. Using this relation in equation (\ref{eq:g_sm_3}), and considering that $|E_F| = \hbar v_Fk_F$, we finally find
 \begin{equation}
\frac{g}{\omega_p} =\frac{\beta}{\left(k_FD\right)^{7/4}},
\end{equation}
where we have collected in $\beta$ all the geometric dimensionless factors
 \begin{equation}
\beta = \frac{\xi_1}{4\sqrt{\xi_2}}\left(\frac{\pi^2 v_{F}}{2\alpha c} \right)^{1/4}\,\Theta_{ijkl}\int_{S} \mathrm{d}^2\mathbf{\hat{r}}\,f_i^{2\omega_p}(\mathbf{\hat{r}})f_j^{\omega_p}(\mathbf{\hat{r}})\hat{\nabla}_k f_l^{\omega_p}(\mathbf{\tilde{r}}).
\end{equation}

Note that, while the scaling of the nonlinearity as $\left(k_FD\right)^{-7/4}$ is a completely general result, the individual terms appearing in $\beta$ are defined in a somewhat arbitrary way. For instance, choosing $D$ to be the length of the short side of a triangle instead of that of the long one, as done for the numerical example in the paper, affects the numerical values of $\xi_1$ and $\xi_2$.

\section{Radiative decay rate computation}
\label{AppDEC}

 There are two ways of obtaining the radiative decay rate of the graphene triangle, one through the extinction cross section, and the other directly through the dipole moment.

 When using the first method, we assume that the polarizability of the graphene triangle can be expressed as a Lorentzian line shape \cite{VST96}
 \begin{equation}
 \label{eq:alpha_w}
 \alpha(\omega)=\frac{6\pi \epsilon_0c^{3}\kappa_{a}}{\omega_{p}^{2}}\frac{1}{\omega_{p}^{2}-\omega^{2}-\mathrm{i}\Gamma'\omega^3/\omega_{p}^{2}},
 \end{equation}
with $\omega_{p}$ being the plasmon frequency, $\Gamma'$ the total decay rate, and $\kappa_{a}$ the radiative contribution to $\Gamma'$. On the other hand, the extinction cross section is directly related to the polarizability by $\sigma^{\rm{ext}}(\omega)= (\omega/c)\,\mathrm{Im}\{\alpha(\omega)\}/\epsilon_0$ \cite{NH06}. The combination of the two former equations at the plasmon frequency results in the final expression for the radiative decay
 \begin{equation}
\hbar\kappa_{a}(\omega_{p})=\frac{\omega_{p}^{2}\sigma^{\mathrm{ext}}(\omega_{p})}{6\pi c^{2}}\hbar\Gamma'.
 \end{equation}
Thus, we can get the value of the radiative decay by obtaining numerically the extinction cross section at the plasmon frequency.

In the second method, the radiative decay is calculated directly from the knowledge of the induced dipole moment $\mathbf{p}$ of the plasmon modes of the triangle, which are numerically computed from $\rm{COMSOL}^{\textregistered}$. Indeed, the radiative decay of a dipole is given by \cite{paper053}
 \begin{equation}
 \hbar\kappa_{a}(\omega_{p})=\frac{\omega_{p}^{3}}{3\pi \varepsilon_{0}c^{3}}|\mathbf{p}_a|^2,
 \end{equation}
 where the dipole moment $\mathbf{p}_a = \int_{S}  \mathrm{d}^2\mathbf{r}\,\rho_a\,\mathbf{r}$ of the mode can be related to the single-plasmon electric field by the continuity equation $\rho_a = (-\mathrm{i}\sigma^{(1)}/\omega)\nabla_{\|}\cdot\tilde{\mathbf{E}}$.

\section{Second harmonic field generated by a hexagonal lattice of nanostructures under strong driving}
\label{AppARR}

Assuming that the lattice is large enough that edge effects can be ignored, the dipoles will respond identically, so that in equation (\ref{array_eq}) of the main paper we get $a_j = a$ and $b_j = b$, and the interaction between them reduces to finding the sum $G = \sum_jG_{ij}$. For a hexagonal lattice this sum consists of a real part that can be approximated as  $\mathrm{Re}[G^{\omega}] \approx(1/4\pi\epsilon_{0})\,5.2/l^3$ and an imaginary contribution whose exact expression is $\mathrm{Im}[G^{\omega}] = (1/4\pi\epsilon_{0})\left[2\pi\omega/cA - 2(\omega/c)^3/3\right]$ \cite{paper182}, where $A = l^2\sqrt{3}/2$ is the area of the unit cell. We now focus on the case of SHG, where the fundamental mode is driven by a strong external input field with frequency around $\omega_p$, while the higher mode is undriven ($E_{2\omega}^{ext}=0$). In the strong field limit, we can replace the operators with numbers.  We can solve for the dipole moments in the frequency domain,
\begin{eqnarray}
\label{b_2}
{b}&=&\left[2\omega - 2\omega_p + \ii\Gamma_b/2 + \frac{p_b^2}{\hbar}\,G^{2\omega_p}\right]^{-1}\,ga^2 \\
&=&-\frac{\hbar}{p_b^2}\,\tilde{\alpha}_b(2\omega)\,ga^2\nonumber
 \end{eqnarray} 
Ignoring the depletion of $a$ due to $b$ (an approximation valid when $a \gg b)$:
\begin{eqnarray}
\label{a_2}
a&=&\left[\omega - \omega_p + \ii\Gamma_a/2 + \frac{p_a^2}{\hbar}G^{\omega_p}\right]^{-1}\,\frac{p_a}{\hbar}\,E^{\mathrm{ext}}_\omega \\
&=& -\frac{E^{\mathrm{ext}}_\omega}{p_a}\,\tilde{\alpha}_a(\omega).\nonumber
 \end{eqnarray} 
In the previous two equations we have defined $\tilde{\alpha}_a(\omega)$ as
\begin{eqnarray}
\tilde{\alpha}_a(\omega)  = \frac{-p_a^2}{\hbar\left(\omega-\omega_p+\ii\Gamma_a/2 + \frac{p_a^2}{\hbar}G^{\omega_p}\right)} = \frac{-p_a^2}{\hbar\left(\tilde{\delta}_a(\omega)+\ii\tilde{\Gamma}_a/2\right)},
 \end{eqnarray} 
and similarly for $\tilde{\alpha}_b(\omega)$. They correspond to the polarizabilities in the proximity of the resonances, modified by the interaction between the structures of the array. Here, $\tilde{\delta}_a(\omega)$ and $\tilde{\Gamma}_a$ are the detuning and the linewidth of the plasmonic mode renormalized respectively by the real and the imaginary part of the Green function.
Inserting equation \,(\ref{a_2}) in (\ref{b_2}) and multiplying by the dipole moment of a single plasmon in the second mode, we find
\begin{equation}
p_{2\omega_p}(\omega) = -\frac{\hbar g}{p_a^2p_b}\,\tilde{\alpha}^2_a(\omega)\tilde{\alpha}_b(\omega)\,\left[{E^{\mathrm{ext}}_\omega}\right]^2 .
\end{equation}

For the lattice constant of interest, the far-field is emitted only in the perpendicular direction \cite{paper182}, and the far-field intensity at frequency $2\omega_p$ under driving at frequency $\omega_p$ can be directly calculated as  
\begin{eqnarray}
\label{intensity}
I_{2\omega_p}^{\rm{far}}&=&\left|\frac{\omega_p}{c\epsilon_0 A}\,p_{2\omega_p}(\omega_p)\right|^2\frac{\epsilon_0c}{2}  \\ &=&\frac{2\hbar^{2}\omega_p^2g^2}{c^3\epsilon_0^3A^2p_a^4p_b^2}\,\left[I^{\mathrm{ext}}_{\omega_p}\right]^2\,| \tilde{\alpha}_a(\omega)|^4|\tilde{\alpha}_b(\omega)|^2 \nonumber\\
&=&\frac{g^2\, {\Gamma_a'}^2\,\Gamma_b'}{8\hbar\omega_p}\frac{[\sigma^{\mathrm{ext}}_a]^2\sigma^{\mathrm{ext}}_b}{A^2}\,\left[I^{\mathrm{ext}}_{\omega_p}\right]^2\,\left[\tilde{\Gamma}_a^2/4 + \tilde{\delta}_a^2\right]^{-2}\left[\tilde{\Gamma}_b^2/4 + \tilde{\delta}_b^2\right]^{-1} ,\nonumber
 \end{eqnarray} 
where $I^{\mathrm{ext}}_{\omega_p} = c\epsilon_0[{E^{\mathrm{ext}}_{\omega_p}}]^2\big{/}2$. 

Now we can calculate the efficiency of conversion using the values obtained for the triangular nanoisland with quality factor $Q=100$, i.e.,\,\,$\kappa_a \approx 2\times10^{-7}\omega_p$, $\kappa_b \approx 5.4\times10^{-8}\omega_p$, and $\Gamma_a' = \Gamma_b' =10^{-2}\omega_p$, $g = 1.25\times10^{-2}\omega_p$. We assume that the lattice period is $l = 50$ nm. The wave vector of the driving light at frequency $\omega_p$ is $k\approx 1$ $\mu\rm{m}^{-1}$. We first calculate the effect of the array on the frequency and the linewidth. The frequency of the first harmonic is redshifted by $6\times10^{-3}\omega_p$, while for the second one the redshift is $2\times10^{-4}\omega_p$. Furthermore, the linewidths $\Gamma_a'$ and $\Gamma_b'$ are increased by  around $17\%$ and $1\%$ respectively, thus we can neglect the effects on the second harmonic. We finally find
\begin{eqnarray}
I_{2\omega_p}^{\rm{far}} \sim 10^{-10}\mathrm{W}^{-1}\mathrm{m}^2\,\left[I^{\mathrm{ext}}_{\omega_p}\right]^2,
 \end{eqnarray} 
an expression valid only when $I_{2\omega_p}^{\rm{far}} \ll I^{\mathrm{ext}}_{\omega_p}$. This calculation indicates that one can observe 1$\%$ of intensity conversion when the driving field has an intensity of about $10^{8}$ $\rm{W\,m^{-2}}$.

\section{Few-photon scattering amplitudes}
\label{AppSCA}

In this section, we present the combined S-matrix and input-output formalism of Ref.\,\cite{FAN10}, which is very helpful to study few-photon scattering amplitudes. We show the equations for a single structure, explaining how to generalize them to the case of the array at the end of the section. We model the incident radiation as a one-dimensional bidirectional continuum, with the two SP modes coupled to it at rates $\kappa_a$ and $\kappa_b$. The Heisenberg equations of motion for the internal mode operators, written in terms of the input operators, are given by \cite{GAD85}
\begin{eqnarray}
\label{eq:heisenberg_2}
\frac{\mathrm{d}a}{\mathrm{d}t}&=&-\ii\left(\omega_p - \ii\Gamma_a'/2\right)a - 2\ii gba^\dagger + F^a_\mathrm{in} + F^a_\mathrm{loss}, \\
\label{d2}
\frac{\mathrm{d} b}{\mathrm{d}t}&=&-\ii\left(2\omega_p - \ii\Gamma_b'/2\right)b - \ii ga^2 + F^b_\mathrm{in} + F^b_\mathrm{loss}.\nonumber
\end{eqnarray}
Here $F^a_\mathrm{in} = -\ii\sqrt{\kappa_{a}/2}\sum_{j=L,R}a^j_\mathrm{in}$ describes coupling from input channels, $F^a_\mathrm{loss} = -\ii \sqrt{\gamma}a_\mathrm{loss}$ describes coupling to loss channels. We have labeled the two directions of propagation for the external light as $L$ and $R$. The relations between input, output and cavity operators are
\begin{eqnarray}
\label{eq:bound}
a^j_\mathrm{out}(t) =  a^j_\mathrm{in}(t) - \ii\sqrt{\frac{\kappa_a}{2}}a(t), \\
\label{d4}
b^j_\mathrm{out}(t) =  b^j_\mathrm{in}(t) - \ii\sqrt{\frac{\kappa_b}{2}}b(t),
\end{eqnarray}
with $j = L,R$. These equations enable one to calculate the properties of the outgoing field based upon the incoming field properties and dynamics of the graphene modes.

The reflection coefficient for a photon propagating in the right direction can be clearly expressed in term of the S-matrix elements
\begin{equation}
\label{eq:r_d}
\bra{p^L_b} S\ket{k^R_b} = \bra{0}b^L_\mathrm{out}(p){b_\mathrm{in}^R}^\dagger(k)\ket{0}.
\end{equation}
Using the Fourier transform of equation (\ref{d4})
\begin{equation}
\label{eq:bound_transf}
b^L_\mathrm{out}(p) = b^L_\mathrm{in}(p) - \ii\sqrt{\frac{\kappa_b}{2}}b(p)
\end{equation}
to express the output operator in equation\,(\ref{eq:r_d}), we get that
\begin{equation}
\label{eq:r_d_2}
\bra{p^L_b} S\ket{k^R_b} = - \ii\sqrt{\frac{\kappa_b}{2}}\bra{0}b(p)\ket{k^R_b} = -\ii\sqrt{\frac{\kappa_b}{4\pi}}\int \mathrm{d}p \bra{0}b(t)\ket{k^R_b}\,e^{\mathrm{i}pt}.
\end{equation}
To determine this matrix element, we use the equation of motion (\ref{d2}) for the operator $b$. Using the commutation relations between the different input operators, $[b^R_\mathrm{in}(k),{b_\mathrm{in}^R}^\dagger(k')] = \delta(k-k')$, $[b^R_\mathrm{in}(k),{b_\mathrm{in}^L}^\dagger(k')] = 0$, etc., we see that only the term with $b^R_\mathrm{in}$ remains:
\begin{eqnarray}
\frac{\mathrm{d}}{\mathrm{d}t} \bra{0}b(t)\ket{k^R_b} =& -\mathrm{i}(2\omega_p - \ii\Gamma_b'/2)\bra{0}b(t)\ket{k^R_b}   \\
 &- \ii g\bra{0}a^2(t)\ket{k^R_b}- \ii\sqrt{\frac{\kappa_b}{2}}\,\bra{0}b^R_\mathrm{in}(t){b_\mathrm{in}^R}^\dagger(k)\ket{0},\nonumber
\end{eqnarray}
where the last term can be immediately calculated using the Fourier transform of the first operator.
Similarly, using the equation of motion for the operator $a^2$, one finds the equation
\begin{equation}
\label{eq:a2_2}
\frac{\mathrm{d}}{\mathrm{d}t}\bra{0}a^2(t)\ket{k^R_b} = -2\mathrm{i}(\omega_p - \ii \Gamma_a')\bra{0}a^2(t)\ket{k^R_b}
 - 2\ii g\bra{0}b(t)\ket{k^R_b}.
\end{equation}
The last two equations constitute a closed system of differential equations in $\bra{0}b(t)\ket{k^R_b}$ and $\bra{0}a^2(t)\ket{k^R_b}$, which can be solved by Fourier transformation. Inserting the results for the former element in equation\,(\ref{eq:r_d_2}), we finally get that \begin{equation}
\label{eq:r_b}
\bra{p^L_b} S\ket{k^R_b} = r_b(k)\delta(p-k),
 \end{equation}
with $r_b(k)$ given by 
\begin{equation}
\label{eq:r_k}
r_b(k) = -\frac{\ii\kappa_b}{2}\frac{\delta_k +\ii\Gamma_a'}{\left[\delta_k +\ii\Gamma_a'\right]\left[\delta_k + \ii\Gamma_b'/2\right] - 2g^2}.
 \end{equation}

With a very similar procedure, the S-matrix elements can be calculated for the down-conversion of a single photon. By a symmetry argument, one readily finds that the DC probability is maximized when the single photon is impinging symmetrically from the modes $L,R$. Provided this condition, we can introduce the symmetric mode $a_\mathrm{in} = 1/\sqrt{2}(a^R_\mathrm{in} + a^L_\mathrm{in})$ (and similarly for $b_\mathrm{in}$ and the output modes) in the equations (\ref{eq:heisenberg_2}-\ref{d4}), and consider a vacuum input in the anti-symmetric mode. The starting point of the calculation is again the expression of the S-matrix element in term of input and output operators:
\begin{equation}
\bra{p_a,q_a} S\ket{k_b} = \bra{0}a_\mathrm{out}(p)a_\mathrm{out}(q)b^\dagger_\mathrm{in}(k)\ket{0}.
 \end{equation}

For what concerns the generalization to an array of $N$ structures, one has only to replace equations (\ref{eq:bound}-\ref{d4}) with 
\begin{equation}
\label{eq:bound_1}
 a_\mathrm{out}(t) =  a_\mathrm{in}(t) - \ii\sqrt{N\kappa_a}A(t).
\end{equation}
\begin{equation}
\label{eq:bound_2}
 b_\mathrm{out}(t) =  b_\mathrm{in}(t) - \ii\sqrt{N\kappa_b}B(t), 
\end{equation}
where $A(t) = N^{-1/2}\sum_i a_i(t)$ and $B(t) = N^{-1/2}\sum_i b_i(t)$ are collective modes, and solve similar differential equations. 

\section*{References}


\end{document}